\documentclass[12pt]{revtex4}
\usepackage{amsfonts}
\usepackage{amsmath}
\usepackage{amssymb}
\usepackage{graphicx}
\usepackage{epsfig}

\begin{document}

\title{On the interactions of the high energy photoelectrons with the
fullerene shell}
\author{E. G. Drukarev $^{1,2}$, M. Ya. Amusia $^{2,3}$}
\affiliation{$^{1}$ B. P. Konstantinov Petersburg Nuclear Physics Institute, Gatchina,
St. Petersburg 188300, Russia\\
$^{2}$The Racah Institute of Physics, The Hebrew University of Jerusalem,
Jerusalem 91904, Israel\\
$^{3}$ A. F.Ioffe Physical-Technical Institute, St. Petersburg 194021, Russia}

\begin{abstract}
The probability that photoionization of the caged atom in an endohedral
system is accompanied by excitation of the fullerene shell is shown to be
close to unity in broad intervals of the photoelectron energies. The result
is obtained by summation of the perturbative series for the interaction
between the photoelectron and the fullerene shell. This result means that
the interaction between the photoelectron ejected from the cage atom and the
fullerene shell cannot be described by a static potential, since inelastic
processes become decisively important.
\end{abstract}

\maketitle


\section{Introduction}

Interaction of the high energy photons with endohedral atoms attracts much
attention nowadays \cite{0c}-\cite{0b}. The inelastic processes in the
fullerene shell (FS) caused by the electrons created in photoionization of
the caged atom is one of the subjects of these studies. The calculations
require understanding of the mechanism of interaction between the
photoelectron and the electrons of the FS. In the present Letter we find the
sum of the cross section of inelastic processes in the FS focusing on the
energy behavior of the ratio
\begin{equation}
r(E)=\frac{\sigma _{A}(E)}{\sigma _{\gamma }(E)}.  \label{5}
\end{equation}%
Here $\sigma _{\gamma }$ and $\sigma _{A}$ are the cross sections of
photoionization of the isolated atom, and that of the caged atom followed by
inelastic processes in the FS (absorbtion cross section); the photoelectron
carries away the energy $E$. We demonstrate that in the broad intervals of
the values of $E$ the ratio $r(E)$ is close to unity. In other words, the
probability that photoionization of the caged atom is followed by an
inelastic process in the FS is close to unity.

This result has consequences for the related problem of studying the wave
function of the outgoing electron in photoionization of the caged atom. We
can not consider the FS as just a source of an external field. Inelastic processes in
the FS are important.

From the first sight the result looks surprising since the two main
mechanisms \cite{1} of excitation and ionization of the FS are connected
with small parameters. In the shake-off (SO) the electronic shell is moved
to an excited state by the sudden change of the effective field caused by
the incoming photon Since FS is far from the caged atom, i.e. the FS radius $%
R$ is big
\begin{equation}
R\gg 1,  \label{2}
\end{equation}%
the SO effects are of the order $1/R^{2}\ll 1$. (We employ the atomic system
of units with $e=m=\hbar =1$).The final state interaction (FSI) between the
photoelectron and the bound ones is determined by its Sommerfeld parameter $%
\xi =1/v$ with $v$ being the relative velocity of the photoelectron and the
bound electrons \cite{1a}. At high energies we can put
\begin{equation}
\xi ^{2}\approx 1/2E\ll 1.  \label{1}
\end{equation}

Note however that while in FSI the electronic shell reacts upon creation of
the hole as a whole, each bound electron is affected by FSI separately.
Hence, the FSI parameter is rather $\xi ^{2}N$ with $N$ being the number of
the bound electrons, while in SO the electronic shell reacts upon creation
of the hole as a whole and the SO effects do not depend on $N$ directly.
Thus, for light atoms $\xi ^{2}N\ll 1$ already at the energies higher than
several hundred eV. However, in the endohedral atoms, which are fullerenes
stuffed with an atom inside, the number of electrons $N$ is much larger and
the FSI become weak at much larger energies. For example, there are $N=360$
electrons in the fullerene $C_{60}$, and $\xi ^{2}N\ll 1$ only for $E\gg 5$
keV.

Fortunately, for the endohedral atoms one can sum the FSI power series in $%
\xi ^{2}N$ without assuming this parameter to be small. This is due to the
large size of these systems -Eq.(\ref{2}). Localization of FS electrons in a
thin layer
\begin{equation}
R\leq r\leq R+\Delta ,  \label{3}
\end{equation}%
enabled us to calculate the sum of power series in $\xi ^{2}N$ in a
model-independent way.

If the photoelectron energy $E$ and thus its momentum $p$ are large enough
so that Eq.(\ref{1}) holds, the first step of the process is the
photoionization of the caged atom, which takes place at the distances of the
order $r\sim 1/p\ll 1$. After that the photoelectron passes the distances of
the order $r\sim R\gg 1$, interacting with the electrons of the FS. Thus,
the amplitude $F_{x}$ of the process with the final state of the electronic
shell $x$ contains the photoionization amplitude $F_{\gamma }(E)$ as a
factor, i.e. \cite{2}
\begin{equation}
F_{x}(E)=F_{\gamma }(E)T_{x}  \label{4}
\end{equation}%
with $T_{x}$ being the amplitude of transition of electrons that belong to
the FS. The accuracy of this equality is of the order $V/E$ with $V$ being
the potential energy of the photoelectron in the field of the FS. In the
lowest order of expansion in powers of $1/R$ we can neglect all the SO
effects and thus $T_{x}$ in Eq.(\ref{4}) is in fact the FSI amplitude.

\section{Lowest order terms}

It is instructive to start with analysis of the lowest order terms. For the
lowest order FSI amplitude of photoionization of caged atom followed by
transition of the FS electrons from the initial state $|\Psi _{0}\rangle $
to an excited state $|\Phi _{x}\rangle $ is given by
\begin{equation}
T_{x}^{(1)}=\langle \Phi _{x}|U_{1}|\Psi _{0}\rangle ,  \label{6}
\end{equation}%
where $U_{1}=\sum_{k}U_{1}(\mathbf{r}^{(k)})$, with $k$ labeling the FS
electron, $U_{1}(\mathbf{r}^{(k)})$ is its interaction with the
photoelectron in the lowest order of the FSI. One can write
\begin{equation}
U_{1}(\mathbf{r}^{(k)})=\frac{1}{c}\int \frac{d^{3}f}{(2\pi )^{3}}G(\mathbf{f%
})g(f)e^{i(\mathbf{fr}^{(k)})},  \label{7}
\end{equation}%
where $c$ is the speed of light \ and
\begin{equation}
G(\mathbf{f})=\frac{-2}{2\mathbf{pf}-i\nu }  \label{7a}
\end{equation}%
is the free electron propagator, in which only the term proportional to the
large momentum $p$ is kept in denominator, $g(f)=4\pi /(f^{2}+\lambda ^{2})$%
, $\lambda \rightarrow 0$. This provides \cite{2}
\begin{equation}
U_{1}=i\xi \Lambda ;\quad \Lambda =\sum_{k}\ln (r^{(k)}-r_{z}^{(k)})\lambda
\label{9}
\end{equation}%
We shall see that the terms containing parameter $\lambda $ form the Coulomb
phase of the $e-e$ scattering and will cancel in the final step.

The second order amplitude is $T_{x}^{(2)}=\langle \Phi _{x}|U_{2}|\Psi
_{0}\rangle $, where
\begin{equation}
U_{2}=\frac{1}{c^{2}}\sum_{k_{1}k_{2}}\int \frac{d^{3}f_{1}}{(2\pi )^{3}}%
\frac{d^{3}f_{2}}{(2\pi )^{3}}G(\mathbf{f_{1}})g(f_{1})G(\mathbf{f_{1}+f_{2}}%
)g(f_{2})e^{i(\mathbf{f}_{\mathbf{1}}\mathbf{r}^{(k_{\mathbf{1}})})}e^{i(%
\mathbf{f}_{2}\mathbf{r}^{(k_{\mathbf{2}})})}.  \label{10}
\end{equation}%
Using Eq.(\ref{7a}) for the Green function $G$ and putting in the integrand $%
G(\mathbf{f_{1}})G(\mathbf{f_{1}+f_{2}})=(G(\mathbf{f_{1}})G(\mathbf{%
f_{1}+f_{2}})+G(\mathbf{f_{2}})G(\mathbf{f_{1}+f_{2}})/2=G(\mathbf{f_{1}})G(%
\mathbf{f_{2}})/2,$ we find that $U_{2}=U_{1}^{2}/2$.

\section{Sum of the power series}
One can see that this
expression can be generalized for the case of arbitrary number $n$ of
interactions between the photoelectron and the FS. Introducing $a_{n}=%
\mathbf{p}(\mathbf{f}_{1}+\mathbf{f}_{2}+..\mathbf{f}_{n})$ we can write
\begin{equation}
\frac{1}{a_{1}}\cdot \frac{1}{a_{2}}...\cdot \frac{1}{a_{n}}=\frac{1}{n!}%
\frac{1}{a_{1}^{n}}.  \label{12}
\end{equation}%
This equation, which can be proved by the induction method was used earlier
for calculation of the radiative corrections in electromagnetic interactions
\cite{3}. Thus, the amplitude
\begin{equation}
F_{x}=F_{\gamma }\langle \Phi _{x}|e^{i\xi \Lambda }|\Psi _{0}\rangle
=F_{\gamma }\langle \Phi _{C}|\Pi _{k}(r^{(k)}-r_{z}^{(k)})^{i\xi
}e^{i\xi\ln {\lambda}}|\Psi _{0}\rangle  \label{12}
\end{equation}%
with $\Lambda $ defined by Eq.(\ref{9}) includes the SO terms and also all
FSI terms with the accuracy $1/R^{2}$. The phase factor $e^{i\xi\ln {\lambda}}$ is canceled by its conjugated
counterpart in the squared amplitude.
Employing Eq.({\ref{3}) we can put $%
r^{(k)}=R$, thus presenting
\begin{equation}
F_{x}=F_{\gamma }T_{x};\quad T_{x}=\langle \Phi _{x}|\Pi
_{k}(1-t^{(k)})^{i\xi }|\Psi _{0}\rangle ,  \label{14}
\end{equation}%
with $t^{(k)}=\mathbf{pr}^{(k)}/pr^{(k)}$.} Here we omitted the constant
factor $(R\lambda)^{i\xi}$,

The absorption cross section can be defined as the difference between the
total cross section and the elastic one . At large $E$ the sum over the
exited states of the FS can be calculated by employing closure
approximation. Thus
\begin{equation}
r(E)=1-|\langle \Psi _{0}|\Pi _{k}(1-t^{(k)})^{i\xi }|\Psi _{0}\rangle |^{2}.
\label{16}
\end{equation}%
This provides
\begin{equation}
r(E)=1-\frac{1}{(1+\xi ^{2})^{N}}=1-e^{-N\ln (1+\xi ^{2})}.  \label{17}
\end{equation}%
If the photon energy is so large that $N\xi ^{4}\ll 1$ ( for $C_{60}$ this
means that $E\gg 800$ eV), we find
\begin{equation}
r(E)=1-e^{-N\xi ^{2}}.  \label{18}
\end{equation}%
In the high energy limit $N\xi ^{2}\ll 1$ the perturbative approach is
valid, and $r(E)\approx N\xi ^{2}$, thus dropping with $E$.

\section{Application to the endohedral atoms $A@C_{\cal N}$}

There are more or less detailed investigated fullerenes with the number of
atoms $\mathcal{N}=20,60,70,80$. There are $N_{v}=4\mathcal{N}$ valence
(collectivized) electrons and $N_{c}=2\mathcal{N}$ core electrons. The
latter can be treated as the $1s$ electrons bound by the carbon nuclei (also
in the field of the valence electrons, the action of which upon $1s$ is
small) with the binding energy $I_{c}\approx 300$ eV . The photoelectron is
known to feel the fullerene potential $V\ll 1$. Thus the condition $\xi
^{2}\ll 1$ enables to use the free propagator presented by Eq.(8). The
binding energies $I_{FS}$ of the valence FS electrons satisfy the condition $%
I_{FS}\ll I_{C}$. Employing of the closure approximation requires that the
photoelectron energy $E$ is large enough to include all important excited
states, i.e. $E$ should be much larger than the energy losses $\bar{%
\varepsilon}$ in the FS. For the energies $\varepsilon $ of the excitation
of the FS, which exceed strongly the FS binding energies $I_{FS}$ (e.g. $%
I_{FS}\approx 7$ eV for $C_{60}$ ) the energy distributions drop as $%
1/\varepsilon ^{2}$. Thus the values of $\bar{\varepsilon}$ for the valence
FS electrons are determined by $I_{FS}\ll \varepsilon \ll E$. They are \cite%
{4}
\begin{equation}
\bar{\varepsilon}=\frac{\xi ^{2}N_{v}}{4R^{2}}\ln \frac{E}{I_{FS}}.
\label{21}
\end{equation}

As follows from Eq. (\ref{21}) the closure approximation can be used for $%
E\gg 50$eV for all the considered fullerenes. At $E\leq I_{C}$ only the
valence electrons can be knocked out by the FSI. In this region $N_{v}\xi
^{2}\geq 3.5$ for $C_{20}$ and is even larger for the fullerenes with larger
$\mathcal{N}$. Thus, $r(E)$ is described by Eq.(\ref{17}) with $N=N_{v}$. At
$E=300$ eV we find $r=0.97$ for $C_{20}$ and $r=0.99997$ for $C_{60}$.
Hence, the cross section $\sigma _{A}$ is very close to the cross section of
photoionization $\sigma _{\gamma }$. At $E\sim I_{c}$ contribution of the
core electrons cannot be calculated by employing the closure approximation.
However Eq.(\ref{17}) provides the lower limit for $r(E)$. For $E\gg I_{c}$
we can employ Eq.(\ref{17}) with $N=N_{v}+N_{c}$. For example, at $E=1$ keV
we find $r=0.80$ for the fullerene $C_{20}$, while $r=0.992$ for $C_{60}$.
The cross section $\sigma _{A}$ is again close to $\sigma _{\gamma }$.

The perturbative behavior requires the energies of the photoelectron to be
very large. We find that $N\xi^2\ll 1$ at $E \gg 1.6 $ keV for the fullerene
$C_{20}$, and at $E \gg 5 $ keV for $C_{60}$. At the energies of dozens of
keV the FSI and SO terms are of the same order and the SO terms should be
included.

\section{Summary}

To summarize, we found the energy dependence of the cross section of
photoionization of the caged atom accompanied by inelastic processes in the
fullerene shell. In the broad intervals of energies this cross section
appeared to be close to that of photoionization of the isolated atom. In
other words, the probability that the FS will be excited during
photoionization of the caged atom is close to unity. This means that in the
related problem of studying the wave function of the outgoing electron in
photoionization of the caged atom we cannot consider the FS as just a source
of external field. Inelastic processes in the FS are important. The results
was obtained by summation of the power series of interaction between the
photoelectron and the FS. The technique of summation can be applied for
investigation of other objects with two scales of distances. It can be
useful for the systems containing a large number of electrons. These can be
heavy atoms or big molecules. The photoionization of the inner shell of a
heavy atom followed by ionization of an outer one can serve as an example.

\section*{Acknowledgments}
The work was supported by the MSTI-RFBR grant 11-02-92484. One of us (EGD)
thanks for hospitality during the visit to the Hebrew University of
Jerusalem.

\end{document}